\documentclass{emulateapj}
\usepackage{apjfonts}
\newcommand{\kel}{\mbox{ K}}
\newcommand{\mkel}{\mbox{ mK}}

\newcommand{\msun}{\mbox{ M$_\odot$}}

\newcommand{\hunits}{\mbox{ km s$^{-1}$ Mpc$^{-1}$}}
\newcommand{\bq}{\begin{equation}}
\newcommand{\eq}{\end{equation}}
\newcommand{\bqa}{\begin{eqnarray}}
\newcommand{\eqa}{\end{eqnarray}}

\newcommand{\bxh}{\bar{x}_H}
\newcommand{\xh}{x_H}

\newcommand{\xixx}{\xi_{xx}}
\newcommand{\xidd}{\xi_{\delta \delta}}
\newcommand{\xixd}{\xi_{x\delta}}

\newcommand{\Pdd}{P_{\delta \delta}}

\newcommand{\mpix}{m_{\rm pix}}

\begin{document}

\title{Statistical Probes of Reionization With 21 cm Tomography}

\author{Steven R.  Furlanetto\altaffilmark{1}, Matias
Zaldarriaga\altaffilmark{2,3}, \& Lars Hernquist\altaffilmark{2}}

\altaffiltext{1} {Division of Physics, Mathematics, \& Astronomy;
  California Institute of Technology; Mail Code 130-33; Pasadena, CA
  91125; sfurlane@tapir.caltech.edu}

\altaffiltext{2} {Harvard-Smithsonian Center for Astrophysics, 60
Garden St., Cambridge, MA 02138}

\altaffiltext{3} {Jefferson Laboratory of Physics, Harvard University, 
Cambridge, MA 02138}

\begin{abstract}

We consider the degree to which ``21 cm tomography'' of the
high-redshift Universe can distinguish different ionization histories.
Using a new analytic model for the size distribution of \ion{H}{2}
regions that associates these ionized bubbles with large-scale galaxy
overdensities, we compute the angular power spectrum and other
statistical properties of the 21 cm brightness temperature during
reionization.  We show that the \ion{H}{2} regions imprint features on
the power spectrum that allow us to separate histories with discrete
bubbles from those with partial uniform ionization (by, for example,
X-rays).  We also show that ``double reionization'' scenarios will
modify the morphology of the bubbles in ways that depend on the
mechanism through which the first generation of sources shuts off.
If, for example, the transition occurs globally at a fixed redshift,
the first generation imprints a persistent feature on the 21 cm power
spectrum.  Finally, we compare our model to one in which voids are
ionized first.  While the power spectra of these two models are
qualitatively similar, we show that the underlying distributions of
neutral hydrogen differ dramatically and suggest that other
statistical tests can distinguish them.  The next generation of
low-frequency radio telescopes will have the sensitivity to
distinguish all of these models and strongly constrain the history and
morphology of reionization.

\end{abstract}
  

\section{Introduction}
\label{intro}

Recently, a great deal of effort -- both observational and theoretical
-- has been focused on understanding the reionization of hydrogen in
the intergalactic medium (IGM) at $z \ga 6$.  Several
independent observational techniques offer complementary views of this
landmark event.  The most straightforward method is to seek extended
troughs of complete Ly$\alpha$ absorption in the spectra of
high-redshift quasars.  Evidence for this \citet{gp} effect
has been found near $z \sim 6.5$ \citep{becker,white03}.
Unfortunately, the optical depth of a fully neutral medium is so high
that current measurements only require a mean neutral fraction $\bxh
\ga 10^{-3}$, and even this limit has been disputed
\citep{songaila04}.  Studies of the rapidly growing ionizing
background \citep{fan} and the Str{\" o}mgren spheres surrounding the
quasars \citep{wyithe04-prox} suggest larger neutral fractions but
depend on detailed modeling.  A second constraint arises because free
electrons scatter cosmic microwave background (CMB) photons, washing
out intrinsic anisotropies and generating a polarization signal \citep{zal97}.
This provides an integral constraint on the reionization history;
recent measurements of the CMB imply that $\bxh$ must have been small
at $z \ga 14$ \citep{kogut03,spergel03}.  Finally, the relatively high
temperature of the Ly$\alpha$ forest at $z \sim 2$--$4$ suggests an
order unity change in $\bxh$ at $z \la 10$
\citep{theuns02-reion,hui03}, although this argument depends on the
characteristics of \ion{He}{2} reionization (e.g.,
\citealt{sokasian02}).

Taken together, these three sets of observations suggest a complex
ionization history extending over a large redshift interval ($\Delta z
\sim 10$).  This is inconsistent with a generic picture of fast
reionization (e.g., \citealt{barkana01}, and references therein).  The
results seem to indicate strong evolution in the sources responsible
for reionization, and a precise measurement of the ionization history
would strongly constrain early structure formation
\citep{wyithe03,cen03,haiman03,sokasian03b}.  Unfortunately,
extracting such detailed information requires new observational
techniques.  One of the most exciting possibilities is ``21 cm
tomography'' of the high-redshift IGM \citep{scott,kumar,mmr}, in
which one maps the distribution of neutral hydrogen on large scales
through its hyperfine transition.  By probing a specific
spectral line, 21 cm surveys measure the time history of reionization,
and unlike the Ly$\alpha$ forest they do not suffer from saturation
problems.  In fact, the signal is sufficiently weak that foregrounds
from the Galaxy and a wide variety of extragalactic objects pose
substantial challenges to these experiments \citep{oh03,dimatteo04}.
Fortunately, the known foregrounds all have smooth continuum spectra,
which should allow frequency cleaning to high accuracy because the 21
cm signal has spectral structure on small scales (\citealt{zald04},
hereafter ZFH04; \citealt{morales03,cooray04}).

However, predicting the 21 cm signals from reionization has proven
difficult.  One method is to use numerical simulations of reionization
\citep{ciardi03,furl-21cmsim,gnedin03}, but computational costs have
so far limited the simulations to subtend (at best) a few resolution
elements of the 21 cm maps.  Analytic models of reionization can
extend to larger scales but require some way to describe the
complexities of structure formation and radiative transfer.  The
simplest approach, in which we assign each galaxy its own \ion{H}{2}
region, does a poor job of matching the large ionized bubbles found in
simulations \citep{ciardi03-sim,sokasian03}.  We have recently
developed a model that associates \ion{H}{2} regions with large-scale
fluctuations in the density field and reproduces the major features
seen in simulations (\citealt{furl04a}, hereafter Paper I).  This
allows us, for the first time, to make quantitative predictions about
the 21 cm signal at reionization.  With high signal-to-noise 21 cm
maps, we can directly measure the distribution of \ion{H}{2} regions
in order to test our model.  However, the signals are sufficiently
weak that such maps will be difficult to make; fortunately,
statistical measurements of the data also contain a great deal of
information about reionization (ZFH04). In this paper, we use the
model of Paper I to predict the 21 cm angular power spectrum for
several reionization scenarios, with an emphasis on how the
measurements help distinguish the crucial features of these different
pictures.  In \S \ref{model} we briefly review our model for
reionization.  We then show how 21 cm measurements can distinguish
different reionization histories in \S \ref{complex} and how they can
distinguish different models of reionization in \S \ref{void}.  We
conclude in \S \ref{disc}.

In our numerical calculations, we assume a $\Lambda$CDM cosmology with
$\Omega_m=0.3$, $\Omega_\Lambda=0.7$, $\Omega_b=0.046$, $H=100 h
\hunits$ (with $h=0.7$), $n=1$, and $\sigma_8=0.9$, consistent with
the most recent measurements \citep{spergel03}.

\section{A Model for Reionization}
\label{model}

Recent numerical simulations (e.g., \citealt{sokasian03}) show that
reionization proceeds ``inside-out'' from high density clusters of
sources to voids, at least when the sources resemble star-forming
galaxies (e.g., Springel \& Hernquist 2003).  We therefore associate
\ion{H}{2} regions with large-scale overdensities.  We assume that a
galaxy of mass $m_{\rm gal}$ can ionize a mass $\zeta m_{\rm gal}$,
where $\zeta$ is a constant that depends on the efficiency of ionizing
photon production, the escape fraction, the star formation efficiency,
and the number of recombinations.  Values of $\zeta \sim 10$--$40$ are
reasonable for normal star formation, but very massive stars can
increase the efficiency by an order of magnitude \citep{bromm-vms}.
The criterion for a region to be ionized by the galaxies contained
inside it is then $f_{\rm coll} > \zeta^{-1}$, where $f_{\rm coll}$ is
the fraction of mass bound to halos above some $m_{\rm min}$.  We will
normally assume that this minimum mass corresponds to a virial
temperature of $10^4 \kel$, at which point hydrogen line cooling
becomes efficient.  In the extended Press-Schechter model
\citep{lacey}, this places a condition on the mean overdensity within
a region of mass $m$,
\bq 
\delta_m \ge \delta_x(m,z) \equiv
\delta_c(z) - \sqrt{2} K(\zeta) [\sigma^2_{\rm min} -
\sigma^2(m)]^{1/2},
\label{eq:deltax}
\eq
where $K(\zeta) = {\rm erf}^{-1}(1 - \zeta^{-1})$, 
$\sigma^2(m)$ is the variance of density fluctuations on the
scale $m$, $\sigma^2_{\rm min}=\sigma^2(m_{\rm min})$ and $\delta_c(z)$
is the critical density for collapse.

Paper I showed how to construct the mass function of \ion{H}{2}
regions from $\delta_x$ in an analogous way to the halo mass function
\citep{press,bond91}.  The barrier in equation (\ref{eq:deltax}) is
well approximated by a linear function, $\delta_x \approx B(m,z) = B_0
+ B_1 \sigma^2(m)$. In that case the mass function has an analytic
expression \citep{sheth98}:
\bq
m \frac{dn}{dm} = \sqrt{\frac{2}{\pi}} \ \frac{\bar{\rho}}{m} \ \left|
  \frac{d \ln \sigma}{d \ln m} \right| \ \frac{B_0}{\sigma(m)} \exp
  \left[ - \frac{B^2(m,z)}{2 \sigma^2(m)} \right],
\label{eq:dndm}
\eq
where $\bar\rho$ is the mean density of the universe.  Equation
(\ref{eq:dndm}) gives the comoving number density of \ion{H}{2}
regions with masses in the range $(m,m+dm)$. The crucial difference
between this formula and the standard Press-Schechter mass function
arises from the fact that $\delta_x$ is a (decreasing) function of
$m$. The barrier is more difficult to cross as one goes to smaller
scales, which gives the bubbles a characteristic size.  In contrast,
the barrier used in constructing the halo mass function is independent
of mass, yielding the usual power law behavior at small masses. The
characteristic scale depends primarily on $\bxh$.

Paper I also showed how to construct the power spectrum of the quantity
$\psi = \xh (1 + \delta)$.  We explicitly computed the correlation
function of the \ion{H}{2} regions $\xixx$ in terms of the mass
function in equation (\ref{eq:dndm}), and we computed the density
correlation function $\xidd$ and cross correlation between density and
ionization fraction $\xi_{x\delta}$ using the halo model (e.g.,
\citealt{cooray02}).  We combined these correlation functions to get
$\xi_\psi$, the correlation function of $\psi$,
\bq
\xi_\psi = \xixx ( 1 + \xidd) + \bxh^2
\xidd + \xi_{x\delta}( 2 \bxh + \xi_{x\delta}).
\label{eq:xipsi}
\eq 
ZFH04 showed how to convert this to the three-dimensional power
spectrum $P_\psi$ and then to the (observable) angular power
spectrum of the 21 cm brightness temperature $\delta T_b$, essentially
by doing a Fourier transform.  Here 
\bqa
\delta T_b \, & \approx \, 23 \,\, \psi & \times
\,\, \left( \frac{T_S - T_{\rm
CMB}}{T_S} \right) \left( \frac{\Omega_b h^2}{0.02} \right) 
\nonumber \\ & & \times \, \, 
\left[ \left(\frac{0.15}{\Omega_m h^2} \right) \, \left(
\frac{1+z}{10} \right) \right]^{1/2} \mkel,
\label{eq:dtb}
\eqa 
with $T_S$ the spin temperature of the gas.  In all of the
calculations here, we assume $T_S \gg T_{\rm CMB}$, which should be
reasonable even relatively early in reionization (see \S 2 of ZFH04).
We further assume perfect frequency resolution in computing the
angular power spectrum; our results therefore do not depend on the
experimental setup.  The effects of finite bandwidth are described in
ZFH04 and Paper I.  We refer the reader to these papers for more
details on our approach.  Note that these methods to construct the
angular power spectrum of $\delta T_b$ can be applied to
any model of the \ion{H}{2} regions.

\section{Complex Reionization}
\label{complex}

Paper I examined the power spectra of standard reionization
scenarios in which all the ionizations are confined to discrete
\ion{H}{2} regions.  We showed that the power spectrum evolves rapidly
throughout reionization, allowing one to reconstruct the ionization
history in such a simple scenario.  Here we extend our model in order
to consider the effects of more complicated reionization processes.

\subsection{A Partially Ionized IGM}
\label{partial}

We first consider how well 21 cm observations can distinguish a model
with all ionizations inside of \ion{H}{2} regions from one with a
uniform component, such as may happen through ionization by X-rays
\citep{oh01b,venkatesan01,ricotti03,madau04} or decaying particles
\citep{sciama82,hansen04}.  Suppose that the IGM has a uniform
ionization fraction $\bar{x}_u(z)$.  On top of this uniform level are
some variations in the neutral fraction owing to isolated \ion{H}{2}
regions.  In this case, the condition $f_{\rm coll} >
\zeta^{-1}(1-\bar{x}_u)$ replaces the barrier in our model: each
galaxy can produce a larger ionized bubble with the same number of
ionizing photons.  However, rather than varying from zero to unity, we
have $0 < \xh < (1 - \bar{x}_u)$.  This damps the fluctuations from
the bubbles, requiring the replacements $\xixx \rightarrow
(1-\bar{x}_u)^2 \xixx$ and $\xixd \rightarrow (1-\bar{x}_u) \xixd$.
Otherwise our formalism is unchanged.

\begin{figure}
\plotone{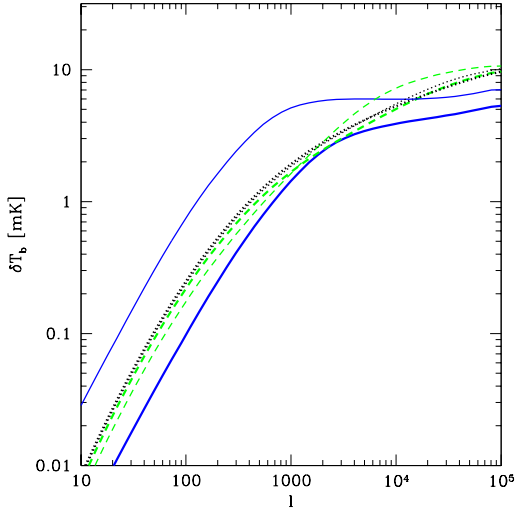}
\caption{The 21 cm fluctuation amplitude as a function of redshift.
  The thick curves have $\zeta=12$ plus a uniformly ionized component,
  while the thin curves have $\zeta=40$.  The ionization fraction
  $\bxh(z)$ is identical in the two models.  The curves have:
  $\bxh=0.96$ (dotted), $\bxh=0.8$ (dashed) and $\bxh=0.26$ (solid).}
\label{fig:xray}
\end{figure}

Figure \ref{fig:xray} contrasts such a case with a ``standard''
history.  The thin curves assume $\zeta=40$ and $\bar{x}_u=0$, while
the thick curves have $\zeta=12$ and set $\bar{x}_u(z)$ to force the
total neutral fraction to match in the two cases.  The uniform
component is thus responsible for $70\%$ of the ionizations.  When
$\bxh \approx 1$, the two curves are nearly identical because the
bubble feature is intrinsically weak.  However, while the bubbles
rapidly imprint a feature on the power spectrum if $\bar{x}_u=0$,
uniform ionization strongly damps the power from the \ion{H}{2}
regions.  As a result, in such a scenario the 21 cm fluctuations trace
those of the density (normalized by $1-\bar{x}_u$), at least until
reionization is almost complete.  Consequently, the amplitude of
$\delta T_b$ is suppressed by $25$--$50\%$ on scales near the
characteristic bubble size, even when $\bxh$ is large.  We find that,
unless reionization is near completion, the bubbles need to provide a
majority of the ionizations in order for them to have an appreciable
effect on the power spectrum.  This provides a clear test for whether
Str{\" o}mgren spheres from ultraviolet photons or smoother ionization
from some other source dominate the morphology of reionization, even
if the survey is unable to make high signal-to-noise maps.

\subsection{Double Reionization}
\label{double}

Recently, a number of theoretical models for reionization have
attempted to reconcile the CMB and quasar data by postulating an early
generation of sources with high ionizing efficiency (most often
because they contain massive, metal free stars) along with a
self-regulation mechanism that switches to normal star formation with
a lower ionizing efficiency
\citep{wyithe03,cen03,haiman03,sokasian03b}.  Such scenarios can lead
to a plateau in the ionized fraction or even ``double'' reionization,
in which ionized phases bracket a mostly neutral period.  In this
section, we explore the signatures such scenarios imprint on the 21 cm
power spectrum through some simple toy models.

We consider an early generation of sources with $(\zeta_1,m_{{\rm
min},1})$ and a later one with $(\zeta_2,m_{{\rm min},2})$.  In the
initial phase only the first generation is important, so the
ionization simply follows the model of Paper I.  We then end this
generation according to one of two conditions.  The first choice shuts
down these sources once the universe reaches some \emph{uniform} level
of ionization at redshift $z_1$.  For example, the early phase could
be caused by X-rays or decaying particles, as described in \S
\ref{partial}.  A better-motivated scenario could be the following.
Suppose that the first generation consisted of massive, metal-free
stars; the natural self-regulation condition halts the formation of
these stars when the metallicity in collapsed objects passes some
threshold \citep{bromm01,mackey03,yoshida03}.  These sources will stop
forming soon after they first appear in any given region (or in other
words soon after the region is ionized for the first time) but they
will continue to form in the remaining neutral regions.  Thus the
first generation will persist until the entire universe has been
ionized.  Our toy model makes the simplification that the first
generation was able to keep the universe ionized until the transition
occurs; we implicitly neglect recombinations within regions that were
ionized before $z_1$.

While the second generation is still rare, recombination in the IGM
will be the dominant effect.  Such an era is obviously easy to
identify with 21 cm tomography because it is accompanied by a
corresponding increase in $\bxh$ and hence in the signal
\citep{furl-21cmsim}.  If the recombination rate were uniform
throughout the IGM, we would have the power spectrum $P_\psi = \bxh^2
\Pdd$.  Because the recombination rate is a function of the density,
we also get fluctuations in the neutral fraction that amplify $\delta
T_b$.  In the limit that the elapsed time is smaller than the
recombination time, we would have $P_{xx} \approx
[1-\bar{x}_i(z)/\bar{x}_i(z_1)]^2 P_{\delta \delta}$, where
$\bar{x}_i$ is the mean ionized fraction.  Unfortunately,
$\xh(\delta,z)$ from recombinations is a nonlinear function, so the
detailed evolution must await numerical simulations.  Given the weak
large-scale fluctuations present at high-redshifts, we do not expect
recombinations to amplify the power by more than a few tens of
percent, and it should not imprint prominent features into the power
spectrum.  (Note that allowing extra recombinations in regions that
were ionized before $z_1$ would increase this amplification factor.)
In our formalism, we can implement this scenario using the results
of \S \ref{partial}, with the extra wrinkle that $\bar{x}_u$ can
decrease with cosmic time.

Another prescription shuts off the first generation before those
sources can fully ionize the universe.  At $z_1$, the
universe has developed a patchwork of \ion{H}{2} regions but
they have not yet overlapped completely.  This choice is appropriate
if the self-regulation mechanism is not local.  For example, the
buildup of an ultraviolet background in the Lyman-Werner bands could
halt H$_2$ cooling and shut off star formation in small halos
\citep{haiman97}, or an X-ray background could heat the universe and
raise the Jeans mass.  In this model, the first generation imprints a
set of ionized bubbles, within which most of the second generation
sources grow (because both appear in the same overdense regions).  The
bubbles grow only slowly until the total number of ionizations from
the second generation become comparable to that of the first; after
this point the evolution approaches the normal behavior.  If we neglect
recombinations within \ion{H}{2} regions, the barrier can be set up as
follows.  Consider a region of mass $m$ at $z_2 < z_1$; as in Paper I
we need the condition for sources inside this region to ionize it.
The number of ionizations is simply the sum of those from both
generations.  Thus we find that the excursion set barrier
$\delta_x(m)$ will be the solution of
\bqa
1 & = & (\zeta_1 - \zeta_2) \, {\rm erfc} \left\{ \frac{\delta_c(z_1) -
  \delta_x(m)}{\sqrt{2[\sigma^2(m_{\rm min,1}) - \sigma^2(m)]}}
\right\} \nonumber \\  & & + \ 
  \zeta_2  \, {\rm erfc} \left\{ \frac{\delta_c(z_2) -
  \delta_x(m)}{\sqrt{2[\sigma^2(m_{\rm min,2}) - \sigma^2(m)]}} \right\}.
\label{eq:double}
\eqa
Here the complementary error functions are the fraction of collapsed
gas above the mass thresholds at the two redshifts.  This is a more
complicated expression than in Paper I, but $\delta_x$ is still
approximately linear in $\sigma^2$.  With the new barrier, the
formalism from Paper I carries over without further modification.  We
could crudely incorporate recombinations into this model by decreasing
$\zeta_1$ as redshift decreases.  This is not exactly correct, because
it would cause the bubbles to shrink rather than recombining at all
radii.  However, this may not be a serious problem because, according
to our prescription, both types of sources will primarily form within
the same highly-biased regions.  Even a relatively small number of
second generation ionizing sources could be enough to halt
recombinations within the bubble unless the recombination time is much
smaller than the Hubble time.  In any case, the main feature of this
model is that the first generation bubbles imprint a particular scale
on the power spectrum.  If recombination is significant (but not
complete) and the second generation sources reside in
partially-ionized bubbles, their own \ion{H}{2} regions will grow
faster than normal and will quickly reach this scale, at which point
their expansion will slow.  If, on the other hand, recombination is
nearly complete in these regions, all evidence of the first generation
disappears and the single generation description of Paper I would
apply.

\begin{figure}
\plotone{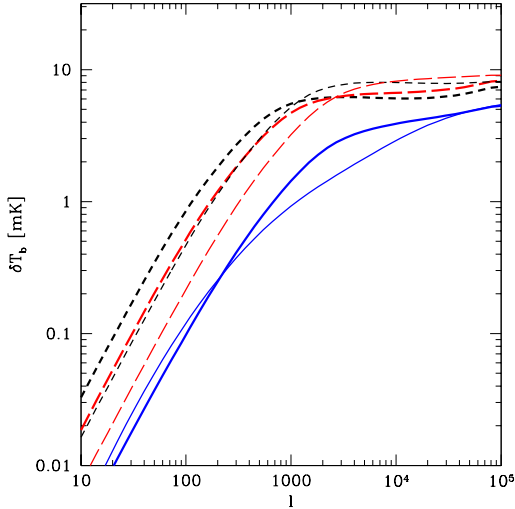}
\caption{The 21 cm power spectrum in three different models of
  ``double reionization.''  The solid curves assume that the universe
  has $\bar{x}_u=0.5$, the long-dashed curves assume $z_1=18$ with
  $\zeta_1=500$, and the short-dashed curves assume $z_1=18$ with
  $\zeta_1=105$ and $m_{\rm min}$ ten times smaller than our default
  value.  In all cases, $\zeta_2=12$.  The thin curves are for $z=16$
  ($\bxh=0.54$) and the thick curves are for $z=12$ ($\bxh=0.73$).}
\label{fig:doublepsi}
\end{figure}

Figure \ref{fig:doublepsi} contrasts the power spectra of three
scenarios for double reionization.  In the solid curves, the first
sources uniformly ionize the IGM to $\bar{x}_u=0.5$ (or, if we neglect
amplification due to recombinations, the IGM has recombined to this
point after a phase of full ionization).  In the other models, we end
the first generation at $z_1=18$ when $\bxh=0.5$; the dotted and
dashed curves correspond to two different sets of parameters for the
first generation of sources.  In all cases the second generation has
$\zeta_2=12$ and ionizes $\sim 4\%$ ($z=16$) and $\sim 23\%$ ($z=12$)
of the IGM.  Note that we have ignored recombinations after $z_1$
in order to isolate the effects of different prescriptions.  The most
obvious feature of the second type of transition is that the bubble
size remains approximately constant with redshift.  This contrasts
sharply with scenarios in which a single set of sources is responsible
for reionization, for which the bubble scale changes rapidly over
$\Delta z \sim 1$ (Paper I).  We suggest that this plateau in the
bubble scale is a clear indication that the transition between
generations occurs rapidly and in a non-local manner.  It is, however,
difficult to distinguish between the two scenarios containing relic
bubbles.  As described in Paper I, this is because the scale of the
\ion{H}{2} regions depends only on the behavior of the large scale
density fluctuations and is fixed almost entirely by $\bxh$.  In the
uniform model, the amplitude of $\delta T_b$ is significantly smaller
and the bubble features are suppressed, just like in \S \ref{partial}.
Note also that in this case the bubble feature grows rapidly, because
the \ion{H}{2} regions expand into a pre-ionized medium.  This stands
in stark contrast to the dashed curves.  While including recombination
could increase the overall amplitude of the uniform model by a small
amount, it would not affect the bubble feature.

\section{Outside-In Reionization}
\label{void}

In Paper I and in \S \ref{complex}, we have examined how the
parameters of the ionizing sources change the 21 cm power spectrum;
however, we have always worked within the model described in \S
\ref{model}.  Many other models for reionization exist, and we now
consider whether the 21 cm signal can discriminate between them.  The
most popular model assumes that the recombination rate controls
ionization \citep{miralda00}.  Thus gas elements below a fixed density
threshold are ionized first, with the threshold increasing as
reionization progresses.  We refer to this as ``outside-in''
reionization because voids are ionized before the dense regions, where
galaxies sit.  A simple way to describe such a process is to assume
that regions below a fixed mean overdensity $\delta_v < 0$ are
ionized.  We can then construct the mass function of the ionized
bubbles through the excursion set formalism; the result is identical
to the \citet{press} halo mass function except that $|\delta_v|$
replaces the collapse threshold.\footnote{In principle, we should
include the ``void-in-cloud'' process described by \citet{sheth03} in
order to remove those voids that have been consumed by collapsed
objects.  However, the regime in which we are interested has
$|\delta_v| \ll \delta_c$, so this process turns out to have no effect
on our results.}  We fix $\delta_v$ by requiring the fraction of mass
in ionized regions with $m > \zeta m_{\rm min}$  (i.e., larger than an
\ion{H}{2} region around the smallest allowed galaxy) to equal $\zeta f_{\rm
coll}$.  This prescription is undoubtedly an oversimplification of the
\citet{miralda00} picture, because it completely neglects the source
locations, the cosmic web, etc.  But it nonetheless serves to
illustrate the importance of our basic assumptions.
The only additional ingredient we need to construct the power spectrum
is the bias of ionized regions.  With the above prescription, we have
\bq
b_v(m,z) = - \left( 1 + \frac{\delta_v^2/\sigma^2 - 1}{|\delta_v|}
\right). 
\label{eq:bias}
\eq
This is the same expression as for the halo bias in the
Press-Schechter model \citep{mo96}, except that the bias is negative
because the number density of voids increases in \emph{underdense}
regions.\footnote{With this prescription, we can have the
cross-correlation between a halo and a void $\xi_{hv} < -1$, which is
unphysical.  In these situations we set $\xi_{hv} = -1$; fortunately,
this has a negligible effect on our results.}  

\begin{figure}
\plotone{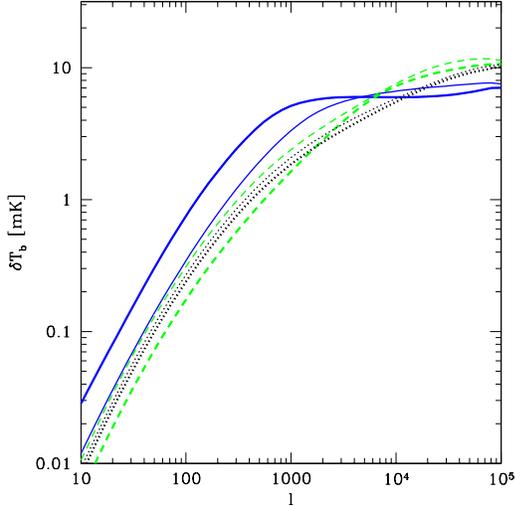}
\caption{The 21cm power spectrum in the inside-out model of Paper I
  (thick lines) and the outside-in model in which voids are ionized
  first (thin lines).  In both cases, $\zeta=40$.  The curves are:
  $\bxh=0.96$ ($z=18$, dotted), $\bxh=0.8$ ($z=15$, dashed),
  and $\bxh=0.26$ ($z=12$, solid).
  }
\label{fig:voidps}
\end{figure}

We compare the outside-in angular power spectrum to the inside-out
model of Paper I in Figure \ref{fig:voidps}.  We have normalized the
two models to have identical $\bxh(z)$.  We see that, until relatively
late in reionization, the fluctuation amplitude and the characteristic
bubble scale are similar in the two models.  This is because both
models associate \ion{H}{2} regions with large-scale fluctuations in
the density field.  The only difference is that the void model barrier
$\delta_v$ is independent of scale but $\delta_x$ increases nearly
linearly with $\sigma^2$ in the inside-out model.  As a result, the
void model has somewhat more ionized gas in smaller regions and a
somewhat smaller characteristic scale; however, these differences may
not be robust given the approximate methods used to construct the size
distribution.  On the other hand, the similarity shows explicitly
that, regardless of the detailed model, the characteristic bubble size
is large (and in the ideal range for 21 cm tomography) so long as the
sizes of \ion{H}{2} regions are determined by fluctuations in the
large-scale density field.  A more fundamental difference between the
two models is that the void model has significantly more large-scale
power, at least until shot noise from the bubbles begins to dominate
at $z \sim 13$.\footnote{As described in Paper I, our model probably
begins to break down at this point because the internal structure of
the \ion{H}{2} regions becomes important.}  By associating ionized
regions with underdensities, the void model \emph{amplifies}
large-scale power.  In contrast, inside-out reionization washes out
large scale power by ionizing overdensities first.

\begin{figure}
\plotone{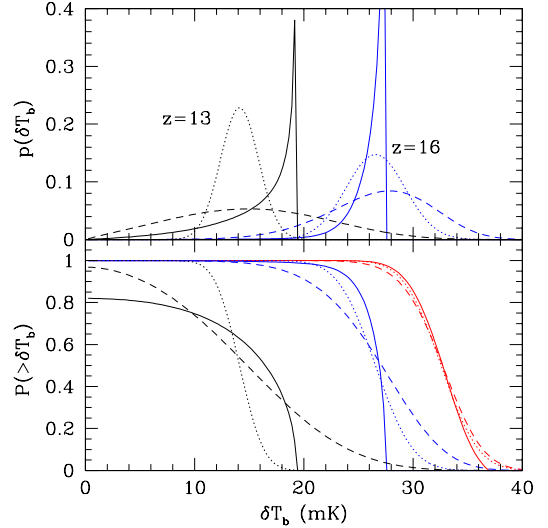}
\caption{The PDF (top panel) and cumulative distribution function
  $P(>\delta T_b)$ (bottom panel) in our three models, assuming
  $\zeta=40$ and $\mpix=10^{13} \msun$.  The solid, dashed, and dotted
  curves are for the inside-out model of Paper I, the outside-in void
  model, and a uniformly ionized IGM, respectively.  In the bottom
  panel, the three sets of curves are for $z=20$ ($\bxh=0.98$), $z=16$
  ($\bxh=0.88$) and $z=13$ ($\bxh=0.5$), from right to left. We only
  show the PDFs for the latter two cases.}
\label{fig:cdf}
\end{figure}

Moreover, as emphasized in Paper I, the power spectrum does not
uniquely determine the statistics of the $\delta T_b$ field.  Although
the density fluctuations are nearly gaussian on the relevant scales,
the \ion{H}{2} regions are not.  We now consider the pixel
distribution function (PDF) in the two models; i.e. the fraction of
spherical ``pixels'' of a given mass $\mpix$ that have brightness
temperature $\delta T_b$.  We showed how to compute this for the
inside-out model in Paper I.\footnote{Note that we do not include
overlap with large bubbles in the results shown here, so these PDFs
cannot be directly compared with observations; see Paper I.}  The same
technique applies to the void model once we modify the excursion set
barrier appropriately.\footnote{In this case $\xh(\delta)$ is a
monotonic function, so the resulting PDF is non-singular.}  We show
some example PDFs in Figure \ref{fig:cdf}; they assume $\mpix=10^{13}
\msun$ or an angular resolution $\sim 2.7'$.  The Figure explicitly
demonstrates that although both reionization models increase the
variance in $\delta T_b$, they do so in qualitatively different ways.
For inside-out reionization, high density regions are highly ionized
and $\delta T_b$ has a maximum value $\delta T_{\rm max}$ on any given
scale.  The pixels tend to cluster strongly around this maximum value
because $\delta T_b$ is a weak function of density in the relevant
range; the PDF has large variance because the mean $\bar{\delta T} <
\delta T_{\rm max}$ thanks to fully ionized regions.  For outside-in
reionization, on the other hand, high-density pixels remain mostly
neutral, while the peak and the low density tail get stretched to
small $\delta T_b$.  The result is not so far from a gaussian, but
$\xh(\delta)$ has dramatically increased the variance.  The bottom
panel shows the cumulative distribution functions $P(> \delta T_b)$
for three redshifts, which is often useful in statistical tests.  Note
that at early times, when $\bxh$ is small, the distributions closely
resemble each other, but significant features appear even when the
ionized fraction is only $\sim 10\%$.  Also, the inside-out model has
$p(\delta T_b=0) \sim 0.2$ at $z=13$, because the characteristic
bubble size is near $\mpix$.  Figure \ref{fig:cdf} shows the necessity
of identifying useful tests for non-gaussianity in the signal.  We
intend to explore such tests in the future.

\section{Discussion}
\label{disc}

We have examined what 21 cm tomography of the high-redshift Universe
can teach us about the ionization history.  Using the reionization
model described in Paper I, we predicted the angular power spectra of
the brightness temperature $\delta T_b$ in several qualitatively
different reionization scenarios.  We have shown that 21 cm
measurements offer a powerful probe of the reionization process, even
without high signal-to-noise maps.  Not only can they easily determine
the timing of reionization, but particular ionization histories have
measurable effects on the power spectrum.  For example, we showed that
the ``bubble'' feature in the power spectrum disappears if a
substantial fraction of the ionizations are due to hard photons that
can travel large distances from their source.  Also, we contrasted the
predictions of two different prescriptions for the transition between
multiple generations of sources.  In one, the transition occurs in an
approximately uniformly ionized medium.  In the other, the transition
occurs while the original \ion{H}{2} regions are still growing.  We
showed that these different prescriptions lead to qualitatively
different power spectra, even long after the first sources have shut
off, because the first generation bubbles can imprint a persistent
feature on the power spectrum.  This could be especially important if
the first phase of reionization ends at redshifts inaccessible to the
next generation radio telescopes, because of either radio frequency
interference or simple noise considerations (note that the background
is approximately proportional to $[1+z]^{2.7}$ for these
measurements).  Finally, we contrasted our ``inside-out'' reionization
model with predictions for an ``outside-in'' model in which voids are
ionized before dense regions.  While the power spectrum of these
models differ in detail, we showed that the underlying distributions
of pixel brightness temperature for a given angular resolution are
dramatically different.  We thus suggest that more sophisticated
statistical measures of the data will be able to strongly constrain
the underlying model of reionization.

Of course, we have made a number of simplifying assumptions in all of
these models.  Most notably, we have ignored recombinations, radiative
transfer, and the anisotropic distribution of sources; we discuss
these issues in more detail in Paper I.  Addressing these
complications in a satisfactory manner will probably require numerical
simulations.  However, none are likely to alter our conclusions: 21 cm
measurements contain an unprecedented wealth of information about the
timing and morphology of reionization.  Our simple model shows that
the qualitative statistical features of the signal can be related to
the source and IGM properties.  The next generation of low-frequency
radio telescopes, such as the Primeval Structure Telescope,\footnote{
See http://astrophysics.phys.cmu.edu/$\sim$jbp for details on PAST.}
the Low Frequency Array,\footnote{ See http://www.lofar.org for
details on LOFAR.} and the Square Kilometer Array,\footnote{ See
http://www.skatelescope.org for details on the SKA.} should be able to
measure these statistical properties (ZFH04) and place strong
constraints on the reionization era.

This work was supported in part by NSF grants ACI AST 99-00877, AST
00-71019, AST 0098606, and PHY 0116590 and NASA ATP grants NAG5-12140
and NAG5-13292 and by the David and Lucille Packard Foundation
Fellowship for Science and Engineering.


\end{document}